\documentclass[11pt,journal,draftcls,onecolumn]{IEEEtran}

\usepackage[dvips]{graphicx}
\usepackage{epsfig}
\usepackage[cmex10]{amsmath}
\usepackage{amssymb}
\usepackage{amsthm}
\usepackage{amsfonts}
\usepackage{bm}
\usepackage{cite}
\usepackage[svgnames]{xcolor} 
\usepackage{pstricks,pst-node,pst-plot,pstricks-add}
\usepackage[tight,footnotesize]{subfigure}
\usepackage{algpseudocode}
\usepackage{algorithm}
\usepackage{enumerate}

\graphicspath{{figs/}}

\input{mohammad.def}

\begin{document}

\title{Private Function Retrieval }

\author{Mahtab Mirmohseni and Mohammad Ali Maddah-Ali%
\thanks{M. Mirmohseni is with the Department of Electrical Engineering, Sharif University of Technology, Tehran, IRAN (e-mail: mirmohseni@sharif.edu).}
\thanks{M. Maddah-Ali is with the Nokia Bell Labs (email: mohammad.maddahali@nokia-bell-labs.com).}
}

\maketitle

\begin{abstract}
The widespread use of cloud computing services raises the question of how one can delegate the processing tasks to the untrusted distributed parties without breeching the privacy of its data and algorithms. Motivated by the algorithm privacy concerns in a distributed computing system, in this paper, we introduce the private function retrieval (PFR) problem, where a user wishes to efficiently retrieve a linear function of $K$ messages from $N$ non-communicating replicated servers while keeping the function hidden from each individual server. The goal is to find a scheme with minimum communication cost. To characterize the fundamental limits of the communication cost, we define the capacity of PFR problem as the size of the message that can be privately retrieved (which is the size of one file) normalized to the required downloaded information bits.
We first show that for the PFR problem with $K$ messages, $N=2$ servers and a linear function with binary coefficients the capacity is $C=\frac{1}{2}\Big(1-\frac{1}{2^K}\Big)^{-1}$.
Interestingly, this is the capacity of retrieving one of $K$ messages from $N=2$ servers while keeping the index of the requested message hidden from each individual server, the problem known as private information retrieval (PIR).
Then, we extend the proposed achievable scheme to the case of arbitrary number of servers and coefficients in the field $GF(q)$ with arbitrary $q$ and obtain $R=\Big(1-\frac{1}{N}\Big)\Big(1+\frac{\frac{1}{N-1}}{(\frac{q^K-1}{q-1})^{N-1}}\Big)$.

\end{abstract}

\begin{IEEEkeywords}
    Private Function Retrieval, private computing, secure distributed storage systems.
\end{IEEEkeywords}

\section{Introduction}
\label{sec:introduction}

Distributed systems are considered as an inevitable solution to store or process large amount of data. However, distributing the computation raises major concerns regarding security and privacy of data and algorithms. This is particularly crucial, if we have to offload the computation and storage tasks to some untrusted, but probably cheaper or more powerful, parties. There is a rich history of study  in the literature for data privacy in distributed environments. However, these days, the algorithm privacy could be even more important than data privacy. Not only the algorithms could be very valuable, but also in some cases parameters of the algorithm could carry lifetime secrets such as biological information of the individuals.  Compared to \emph{data privacy}, our understanding of the fundamental limits of \emph{algorithm privacy} is very limited.

Motivated by this, we introduce the private function retrieval (PFR) problem, where a set of servers with access to a database is connected to a user. The user wishes to compute a function of the files while keeping the function private from each individual server. The goal is to characterize the fundamental limits of the communication cost (between the user and the servers) needed in order to privately compute the function.

Recently, there has been intense interest in characterizing the fundamental performance limits of distributed computing systems from an information theoretic perspective. Among these, we can name the distributed storage systems \cite{dimakis10}, distributed cache networks \cite{maddah14}, private information retrieval (PIR) \cite{chor95,shah14,Sun17}, and distributed computing~\cite{li17,yu17}. In all of these cases, information theoretic ideas and tools have been found useful to provide a fundamental, and often very different, understanding on how to run the system efficiently. In this work, our goal is to characterize the fundamental limits of PFR from an information theoretic perspective.

To be precise, in this paper, we consider a system including one user connected to $N$ non-colluding servers, each storing a database of $K$ equal-size files, $W_1, \dots, W_K$. The user wishes to compute a linear combination of these $K$ files by downloading enough equations from the $N$ servers. While retrieving the linear combination, the user wishes to keep the coefficients hidden from each individual server. This means that each server must be equally likely uncertain about which combination is requested by the user. The goal is to minimize the required downloading load to retrieve the result of computation privately.

The PFR problem can be considered as an extension of the PIR problem, where the user is interested in one of $K$ files. The PIR problem has been introduced in \cite{chor95} and its capacity in the basic setup has been characterized recently in~\cite{Sun17}. Several extensions of PIR problem has been studied in literature, including the PIR with colluding servers~\cite{sun16colluding,banawan17colluding}, the PIR with coded servers~\cite{banawan16coded}, and the symmetric PIR~\cite{sun16symmetric,wang17symmetric}.

To address the problem of PFR, we first focus on the cases where the coefficients are from a binary field. For this case, we find the optimal scheme for two servers ($N=2$) and any arbitrary number of files, $K$.  In particular, we show that the capacity of this case is $\frac{1}{2}\Big(1-\frac{1}{2^K}\Big)^{-1}$. Interestingly, this is equal to the capacity of the PIR with two servers and arbitrary number of files, $K$.
We extend this scheme, and propose an achievable solution for the general setup with $N$ servers, $K$ files and coefficients from a general valid field.

The capacity of PFR problem has been studied in a parallel and independent work~\cite{sun17PFR}. In~\cite{sun17PFR}, the capacity of PFR has been characterized for a system with two servers ($N=2$), two messages ($K=2$) and arbitrary linear combination. In this paper, we characterize the capacity of PFR for a system with $N=2$ servers, an arbitrary number of files, and binary coefficients. The achievable schemes proposed by two papers are very different.

The remainder of this paper is organized as follows.
Section~\ref{sec:problem} formally introduces our information-theoretic formulation of the PFR problem.  Section~\ref{sec:main} presents main results. Sections~\ref{sec:scheme_binary} and \ref{sec:scheme_gen} contain proofs.

\section{Problem Setting}
\label{sec:problem}

We consider a system, including a user connected to $N$ non-colluding servers, each stores an identical copy of a database.
The database incudes $K$ files $W_1, \dots, W_K$, where each file $W_k$ has $L$ equal-size segments (or so-called layered) $W_k[1],\dots,W_k[L]$ for $k=1,\dots,K$, i.e., $W_k=\{W_k[t]\}_{t=1}^{L}$. Each segment $W_k[t]$, $t\in [1:L]$, $k\in [1:K]$ is chosen independently and uniformly at random from the finite field $GF(p^F)$, for some $F\in\N$ and prime number $p$. The database is shown as $\left\{ \W[t] \right\}_{t=1}^L$, where $\W[t]=[W_1[t], \dots, W_K[t]]^T\in (GF(p^F))^K$.

The user is interested in a specific linear function of $\W[t]$, represented as
\begin{equation}
\label{eq:linear}
   \{\vv^T\W[t]\}_{t=1}^L,
\end{equation}
where $\vv$ is an $K$--dimensional non-zero vector, with entries from the finite field $GF(q)$, for some integer $q$.  We assume that $GF(q)$ is a sub-field of $GF(p^F)$,  thus $q=p^M$, for some integer $M$, where $M|F$. Therefore, the operations in \eqref{eq:linear} are well-defined over $GF(p^F)$. Excluding the parallel vectors, there are $\frac{q^K-1}{q-1}$ distinct options for vector $\vv$, denoted by  $\vv(1),\dots,\vv(\frac{q^K-1}{q-1})$. We use a short-hand notation $M(i,t)=\vv^T(i)\W[t]$.

Note that in the PFR problem with binary coefficients, we set $p=2$ and $M=1$ that yields $q=2$.

Assume that the user chooses $\vv=\vv(\theta)$ for some $\theta \in [1:\frac{q^K-1}{q-1}]$, thus the user wishes to compute $\{{M}(\theta,t)\}_{t=1}^L=\{\vv^T(\theta)\W[t]\}_{t=1}^L$ by downloading some equations from the servers. So, the user sends $N$ queries $Q_{\vv(\theta)}^{(1)},\dots,Q_{\vv(\theta)}^{(N)}$, to server 1 to $N$ respectively,  where $Q_{\vv(\theta)}^{(n)}$ is the query sent by the user to the $n$-th server in order to retrieve $\{M(\theta,t)\}_{t=1}^L=\{\vv^T(\theta)\W[t]\}_{t=1}^L$. Since the queries are independent of the messages, we have
\begin{equation}\label{eq:query}
    I(\{ \W[t] \}_{t=1}^L;Q_{\vv(i)}^{(1)},\dots,Q_{\vv(i)}^{(N)})=0,
\end{equation}
for all $i\in\left[1:\frac{q^K-1}{q-1}\right]$.

In response to $Q_{\vv(\theta)}^{(n)}$, the $n$-th server computes an answer $A_{\vv(\theta)}^{(n)}$ as a function of its database and the received query, thus
\begin{equation}\label{eq:answer}
    H(A_{\vv(\theta)}^{(n)}|Q_{\vv(\theta)}^{(n)},\{ \W[t] \}_{t=1}^L)=0.
\end{equation}

Let $\max\limits_{\theta} \sum\limits_{n=1}^N|A_{\mathbf{v(\theta)}}^{(n)}|=QF$, for some integer $Q$, represent the download cost in $p$-ary units.

While retrieving $\{M(\theta,t)\}_{t=1}^L=\{\vv^T(\theta)\W[t]\}_{t=1}^L$ from $A_{\vv(\theta)}^{(1)},\dots,A_{\vv(\theta)}^{(N)}$ and $Q_{\vv(\theta)}^{(1)},\dots,Q_{\vv(\theta)}^{(N)}$, the user must keep the index $\theta$, (or equivalently the vector $\vv(\theta)$) hidden from each individual server. To satisfy the privacy constraint, the $\frac{q^K-1}{q-1}$ query-answer function $(Q_{\vv(i)}^{(n)},A_{\vv(i)}^{(n)}),i=1,\dots,\frac{q^K-1}{q-1}$ must be identically distributed in each server $n=1,\dots,N$. That is
\begin{equation}\label{eq:privacy}
    (Q_{\vv(1)}^{(n)},A_{\vv(1)}^{(n)},\{ \W[t] \}_{t=1}^L)\sim(Q_{\vv(\theta)}^{(n)},A_{\vv(\theta)}^{(n)},\{ \W[t] \}_{t=1}^L)
\end{equation}
for each $\theta\in\{1,\dots,\frac{q^K-1}{q-1}\}$ and $n=1,\dots,N$.

An $(L,Q)$ PFR scheme consists of $N(\frac{q^K-1}{q-1})$ query-answer function $(Q_{\vv(\theta)}^{(n)},A_{\vv(\theta)}^{(n)})$ for $i=1,\dots,\frac{q^K-1}{q-1}$ and $n=1,\dots,N$; and $\frac{q^K-1}{q-1}$ decoding functions that map $\{Q_{\vv(\theta)}^{(n)},A_{\vv(\theta)}^{(n)}\}_{n=1}^N$ to $\{ \hat{M}(\theta,t) \}_{t=1}^L$  as the estimate of $\{{M}(\theta,t)\}_{t=1}^L$ for $i=1,\dots,\frac{q^K-1}{q-1}$ with probability of error
$$P_{e,L}=\max_{\theta}\Pr\left(\{ M(\theta,t) \}_{t=1}^L \neq \{\hat{M}(\theta,t) \}_{t=1}^L\right),$$
while the privacy constraint \eqref{eq:privacy} is satisfied.

The rate of this code is defined as
\begin{equation}\label{eq:rate}
    R=\frac{L}{Q}.
\end{equation}

\begin{definition}
A rate $R$ is achievable if there exists a sequence of $(L,Q)$ PFR schemes where $P_{e,L}\rightarrow0$ as $L\rightarrow\infty$.
The capacity of PFR problem is defined as
\begin{equation*}
    C \defeq \sup \big\{R: R \text{ is achievable} \big\}.
\end{equation*}
\end{definition}

Thus from the Fano's inequality, the correctness condition, i.e., $P_{e,L}\rightarrow0$, implies that
\begin{equation}\label{eq:correctness}
    \frac{1}{L}H(\{M(i,t)\}_{t=1}^L|Q_{\vv(i)}^{(1)},A_{\vv(i)}^{(1)},\dots,Q_{\vv(i)}^{(N)},A_{\vv(i)}^{(N)})=o(L),
\end{equation}
where from the Landau notation, we have $f(n)=o(g(n))$ if $\lim\limits_{n\rightarrow \infty}\frac{f(n)}{g(n)}\rightarrow 0$.

\section{Main Results}\label{sec:main}

The first theorem presents the capacity of the PFR problem with binary coefficients ($q=2$) when $N=2$ servers are available with $K$ arbitrary messages.

\begin{theorem}\label{thm:binary_capacity}
For the  PFR problem, with K messages and $N=2$ servers and binary coefficients, the capacity is
\begin{equation}\label{eq:capacity_binary}
    C= \frac{1}{2}\left(1-\frac{1}{2^K}\right)^{-1}.
\end{equation}
\end{theorem}

\begin{remark}
Recall that the user needs the results of $\{\vv^T \W[t] \}_{t=1}^{L}$  for some integer $L$ and  for some $\vv\in (GF(2))^K\setminus\{\mathbf{0}\}$. Clearly $\vv$ has $2^K-1$ options  listed in a set $\mathcal{V}=\{\vv(1), \ldots, \vv(2^K-1)\}$. Therefore, the goal is to design an achievable scheme which has two properties: (1) \emph{correctness}, meaning the user can decode what is asked for, (ii) \emph{privacy} meaning that for every single server, all members of  $\mathcal{V}$ are equiprobable, independent of the real $\vv$. The above theorem states that minimum communication load, normalized to the size of a file, to guarantee  both privacy and correctness is $\frac{1}{2}\left(1-\frac{1}{2^K}\right)^{-1}$.
\end{remark}

\begin{remark}
In the proposed achievable scheme, the set of requests to each server is symmetric with respect to all vectors in $\mathcal{V}=\{\vv(1), \ldots, \vv(2^K-1)\}$, thus the privacy is guaranteed.  However, the requests of two servers are coupled to exploit two opportunities. In the first opportunity, every requests from a server, except a few, has a counterpart request from the other server, such that these two together can reveal  $\vv^T \W[t]$ for some $t$. This justifies the factor of $\frac{1}{2}$ in \eqref{eq:capacity_binary}.
In the second opportunity, in some cases, a request from one server directly reveals a value of $\vv^T \W[t]$ for some $t$. This has been reflected in the factor $\left(1-\frac{1}{2^K}\right)^{-1}$ in~\eqref{eq:capacity_binary}. These two opportunities are exploited together efficiently such that not only the correctness and privacy have been guaranteed, but also the scheme achieves the optimal bound.
\end{remark}

\begin{remark}
We note that for this case, the user asks for the results of $\{\vv^T \W[t] \}_{t=1}^{L}$, where $\vv$ has  $2^K-1$ options listed in a set $\mathcal{V}=\{\vv(1), \ldots, \vv(2^K-1)\}$. Therefore, the user wants to hide its requested combination $\{\vv^T \W[t] \}_{t=1}^{L}$ among $2^K-1$ (virtual) files, namely
$\{\vv^T(1) \W[t] \}_{t=1}^{L}, \ldots, \{\vv^T(2^K-1) \W[t] \}_{t=1}^{L}$. Apparently these virtual files are not linearly independent. One solution for this problem is to ignore this dependency, and to consider a PIR problem with $2^K-1$ virtual files. That approach achieves the rate of $\frac{1}{2}\left(1-\frac{1}{2^{2^K-1}}\right)^{-1}$ (see~\cite{Sun17} for the rate of PIR). However, here, the surprising fact is that the proposed scheme achieves the rate of $\frac{1}{2}\left(1-\frac{1}{2^K}\right)^{-1}$, as if there are only $K$ options for $\vv$. This is done by efficiently exploiting the linear dependency of vectors in $\{\vv(1), \ldots, \vv(2^K-1)\}$.
\end{remark}

\begin{remark}
The PFR problem with binary coefficients reduces to the PIR problem if we restrict the possible coefficient vectors $\vv$ to those with \emph{unit} Hamming weight. Thus, the converse of PIR  problem in\cite[Theorem~1]{Sun17} with $N=2$ is valid for the PFR problem with binary coefficients. The proposed achievable scheme detailed in~\ref{sec:scheme_binary} meets this converse.
\end{remark}

The next lemma extends the achievable scheme of Theorem~\ref{thm:binary_capacity} to the case of arbitrary number of servers and arbitrary $GF(q)$ field for the coefficient vectors $\vv(i)$.

\begin{lemma}\label{lemma:ach_gen}
For the PFR problem with $N$ servers, $K$ messages, and the coefficient vectors $\vv\in(GF(q))^K\setminus\{\mathbf{0}\}$, if $q\geq N$, the following rate is achievable.

  \begin{equation}\label{eq:ach_gen2}
    R=\Big(1-\frac{1}{N}\Big) \cdot \Big(1+\frac{\frac{1}{N-1}}{(\frac{q^K-1}{q-1})^{N-1}}\Big)
\end{equation}

\end{lemma}

\begin{remark}
In this case, the user needs the results of $\{\vv^T \W[t] \}_{t=1}^{L}$  for some integer $L$ and  for some $\vv\in (GF(q))^K \setminus \{\mathbf{0}\}$. Eliminating parallel vectors in $(GF(q))^K \setminus \{\mathbf{0}\}$, there are $\frac{q^K-1}{q-1}$ options for $\vv$, listed in the set $\mathcal{V}=\{\vv(1), \ldots, \vv(\frac{q^K-1}{q-1})\}$.
If we treat each of $\{\vv^T(i)\W[t] \}_{t=1}^{L}$, for $i=1,\ldots, \frac{q^K-1}{q-1}$ as a virtual file, and apply the PIR scheme for these virtual files, we achieve the rate of
$$\left(1-\frac{1}{N}\right) \cdot \left( 1-N^{-\frac{q^K-1}{q-1} }   \right)^{-1}.$$

One can verify that the proposed scheme strictly outperforms the PIR-based scheme.
\end{remark}

\begin{corollary}
For the PFR problem with $K$ messages and  the coefficient vectors $\vv\in(GF(q))^K\setminus \{\mathbf{0}\}$, with $N\rightarrow \infty$ servers,
the capacity is equal to
\begin{equation}\label{eq:ach_gen}
    C=1-\frac{1}{N} .
    \end{equation}
\end{corollary}
The above corollary derives directly from Lemma~\ref{lemma:ach_gen}. This rate meets the PIR converse.

\section{PFR Scheme with binary coefficients (Achievability Proof of Theorem~\ref{thm:binary_capacity})}\label{sec:scheme_binary}
In this section, we present the achievable scheme for the PFR problem with two servers ($N=2$) and arbitrary number of messages $K$, where the coefficients are from the binary field.

The proposed scheme guarantees the privacy by keeping the requests to one server symmetric with respect to all $\vv(i) \in \mathcal{V}$. However, the requests to both servers are coupled in a certain way. In most of the cases, each request to one server has a counterpart in the set of requests from the other server. These two together reveals $\vv^T(\theta)\W[t]$ for some $t$. Some other requests directly reveals $\vv^T(\theta) \W[t]$ for some $t$ without any recombining with other server.

Let $L=2^{K+1}$ and define $\mathcal{V}\defeq(GF(2))^K\setminus \{\mathbf{0}\}$. Also, consider $\pi$ as a random permutation of the set $\{1,2,\dots,L\}$. The user generates this permutation, uniformly at random, among all permutations, and keeps it private from the servers. Apply this random permutation to reorder the messages. In particular, reorder the message vectors to get $\tilde{\W}[t]\defeq \W[\pi(t)]$ for $t\in\{1,\dots,L\}$. Without loss of generality assume that the user is interested in retrieving  $\{\vv(\theta)\W[t]\}_{t=1}^L$, for some $\theta \in \{1, \ldots, 2^K-1\}$.

\textbf{Phase 1}:
\begin{enumerate}[(i)]
  \item User asks server~1 to send back
  \begin{equation}\label{eq:phase1-1}
    M(i,i)=\vv^T(i)\tilde{\W}[i], \quad i=1,\dots,2^K-1.
  \end{equation}

  \item User asks server~2 to send back
  \begin{equation}\label{eq:phase1-2}
    M(i,2^K-1+i)=\vv^T(i)\tilde{\W}[2^K-1+i], \quad i=1,\dots,2^K-1.
  \end{equation}
\end{enumerate}

\textbf{Phase 2}:
\begin{enumerate}[(i)]
  \item User asks server~1 to send back
  \begin{equation}\label{eq:phase2-11}
    M(\theta,2(2^K-1)+1)=\vv^T(\theta)\tilde{\W}[2(2^K-1)+1],
  \end{equation}
  and also
  \begin{equation}\label{eq:phase2-12}
    (\vv^T(\theta)-\vv^T(i))\tilde{\W}[2^K-1+i], \quad i=1,\dots,2^K-1, \ i \neq \theta.
  \end{equation}

  \item User asks server~2 to send back
  \begin{equation}\label{eq:phase2-21}
    M(\theta,2(2^K-1)+2)=\vv^T(\theta)\tilde{\W}[2(2^K-1)+2],
  \end{equation}
  and also
  \begin{equation}\label{eq:phase2-22}
    (\vv^T(\theta)-\vv^T(i))\tilde{\W}[i], \quad i=1,\dots,2^K-1, \ i \neq \theta.
  \end{equation}

\end{enumerate}
It is important to note  that the above requests will be send to the servers in a random order.

The requests and answers from server~1 and server~2 for $\theta=1$ are shown in Table~\ref{tbl:scheme_ser1} and Table~\ref{tbl:scheme_ser2}, respectively.

\begin{table*}
\renewcommand{\arraystretch}{1.3}
\setlength\tabcolsep{3pt}
\caption{Requests from server~1 for $\theta=1$}
\label{tbl:scheme_ser1} \centering
\begin{tabular}{|c||cccc||cccc||cc|}
\hline
& $\tilde{\W}[1]$ & $\tilde{\W}[2]$ &\dots& $\tilde{\W}[2^K-1]$& $\tilde{\W}[2^K]$ & $\tilde{\W}[2^K+1]$& \dots& $\tilde{\W}[2^{K+1}-2]$ & $\tilde{\W}[2^{K+1}-1]$ & $\tilde{\W}[2^{K+1}]$\\
\hline
\hline
$\vv(1)$& $\vv^T(1)$ &    &&   &   &    & &      &$\vv^T(1)$          &       \\
$\vv(2)$&  &  $\vv^T(2)$  &&   &   &  $ (\vv^T(1)-\vv^T(2))$  & &      &          &       \\
\dots&  &    &\dots&   &   &    & \dots&      &          &       \\
$\vv(2^K-1)$&  &    &   &  $\vv^T(2^K-1)$ &   &    & &    $ (\vv^T(1)-\vv^T(2^K-1))$   &          &       \\
\hline
\end{tabular}
\end{table*}

\begin{table*}
\renewcommand{\arraystretch}{1.3}
\setlength\tabcolsep{2pt}
\caption{Requests from server~2 for $\theta=1$}
\label{tbl:scheme_ser2} \centering
\begin{tabular}{|c||cccc||cccc||cc|}
\hline
& $\tilde{\W}[1]$ & $\tilde{\W}[2]$ &\dots& $\tilde{\W}[2^K-1]$& $\tilde{\W}[2^K]$ & $\tilde{\W}[2^K+1]$& \dots& $\tilde{\W}[2^{K+1}-2]$ & $\tilde{\W}[2^{K+1}-1]$ & $\tilde{\W}[2^{K+1}]$\\
\hline
\hline
$\vv(1)$&  &    &&   &$\vv^T(1)$   &    & &      &         &  $\vv^T(1)$      \\
$\vv(2)$&  &  $ (\vv^T(1)-\vv^T(2))$  &&   &   &  $ \vv^T(2)$  & &      &          &       \\
\dots&  &    &\dots&   &   &    & \dots&      &          &       \\
$\vv(2^K-1)$&  &    &   &  $ (\vv^T(1)-\vv^T(2^K-1))$ &   &    & &    $ \vv^T(2^K-1)$   &          &       \\
\hline
\end{tabular}
\end{table*}

\subsection{Proof of correctness}\label{subsec:correct}
To prove the correctness, we show that the user can recover $\{\vv^T(\theta)\W[t]\}_{t=1}^L$ from the combinations \eqref{eq:phase1-1}-\eqref{eq:phase2-22}, received from both servers, while the rate of the scheme is equal to \eqref{eq:capacity_binary}.

We remind that $L=2^{K+1}$, and thus $2^{K+1}$ combinations must be derived from the available equations at the user. $\vv^T(\theta)\tilde{\W}[\theta]$ is given in \eqref{eq:phase1-1}. To obtain $\vv^T(\theta)\tilde{\W}[t]$ for all $t \in \{1, 2,\dots,2^K-1\}\setminus \{\theta \}$, the user combines \eqref{eq:phase1-1} and \eqref{eq:phase2-22}. Similarly, $\vv^T(\theta)\tilde{\W}[2^K-1+\theta]$ is given in \eqref{eq:phase1-2} and $\vv^T(\theta)\tilde{\W}[t]$ for all $t \in \{ 2^K, \dots,2^{K+1}-2 \} \setminus \{ 2^K-1+\theta \}$ can be obtained by combining \eqref{eq:phase1-2} and \eqref{eq:phase2-12}. Finally, $\vv^T(\theta)\tilde{\W}[2^{K+1}-1]$ and $\vv^T(\theta)\tilde{\W}[2^{K+1}]$ are given in \eqref{eq:phase2-11} and \eqref{eq:phase2-21}, respectively.

The total number of downloads is
\begin{equation*}
    Q=2(2^{K+1}-2)=4(2^K-1)
\end{equation*}
and so the rate of the code is
\begin{equation*}
    R=\frac{L}{Q}=\frac{2^{K+1} }{4(2^K-1)}=\frac{1}{2}\left(1-\frac{1}{2^K} \right)^{-1}.
\end{equation*}

\subsection{Proof of privacy}\label{subsec:privacy}
Our privacy proof is based on the fact that we preserve the equal number of requests for any possible coefficient vector in addition to using a random permutation over the message layers. Furthermore, we send the requests to each server in a random order.

First, consider server~1 with its requests \eqref{eq:phase1-1}, \eqref{eq:phase2-11} and \eqref{eq:phase2-12}. As seen, server~1 only observes that the user requests a linear combination for $2^{K+1}-1$ layers of messages, while two layers are left out. The indices of these layers do not leak any information about the requested combinations vector $\vv(\theta)$, thanks to the random permutation of the message layers. Now, let's check the requested coefficient vectors in  \eqref{eq:phase1-1}, \eqref{eq:phase2-11} and \eqref{eq:phase2-12}. We note that the set $\{\{\vv(\theta)-\vv(i)\}_{i\in [1:2^K-1]\setminus\{\theta\}},\vv(\theta)\}$ is equal to the set $\mathcal{V}=\{\vv(1),\dots,\vv(2^K-1)\}$. This means that each possible coefficient vector $\vv\in\mathcal{V}$ is requested exactly twice, and in a random order, and thus no information can be obtained by server~1. In fact, it can be easily shown that for any coefficient vector $\vv(j),j=1,\dots,2^K-1$, there is a permutation $\pi$ of the set $\{1,2,\dots,L\}$ that maps the requests of $\{\vv^T(j) \W[t]\}_{t=1}^L$ from one server to the requests of $\{\vv^T(\theta) \W[t]\}_{t=1}^L$ from the same server. The privacy condition at server~2 is guaranteed similarly.

\section{General PFR Scheme (Proof of Lemma~\ref{lemma:ach_gen})}\label{sec:scheme_gen}

In this section,  we present the general achievable scheme for the PFR problem with $N$ servers, $K$ messages and the linear combinations over $GF(q)$. At first, we define  some notations. We define
\begin{align}
\mathcal{V}=(GF(q))^K\setminus \{\mathbf{0}\}
\end{align}
as the set of all options for $\vv(\theta)$. In addition, for each $\vv(\theta)\in\mathcal{V}$, we define
\begin{align}
\mathcal{V}^{(\theta)}=\left\{
\beta \vv(\theta),
 \beta \in GF(q)\setminus\{0\}  \right\},
\end{align}
as the set of all parallel and non-zero vectors  to $\vv(\theta)$. We also define $\mathcal{V}_N\defeq\mathcal{V}^{N-1}$ as a set of all $(N-1)$-tuples of vectors $(\vv(i_1),\dots,\vv(i_{N-1}))$ with each element from $\mathcal{V}$ (all possible $\vv$ vectors):
\begin{equation*}
    \mathcal{V}_N=\{(\vv(i_1),\dots,\vv(i_{N-1})): \vv(i_j)\in\mathcal{V},j=1,\dots,N-1\}.
\end{equation*}
Note that $ | \mathcal{V}_N|=(q^K-1)^{N-1}$.

Moreover, we define $\mathcal{V}^{(\theta)}_N$ as a set of all $(N-1)$-tuples of vectors with each element from $\mathcal{V}^{(\theta)}$:
\begin{equation*}
  \mathcal{V}^{(\theta)}_N=\{(\vv(i_1),\dots,\vv(i_{N-1})): \vv(i_j)\in\mathcal{V}^{(\theta)},j=1,\dots,N-1\}.
\end{equation*}
Apparently, $\mathcal{V}^{(\theta)}_N \subset \mathcal{V}_N$. Note that $|\mathcal{V}_N^{(\theta)}|=(q-1)^{N-1}$.

Now we are ready to detail the proposed scheme in three steps.

\textbf{Step 1:} Consider $L$ layers of messages. The user generates  a random permutation $\pi$ of the set $\{1,2,\dots,L\}$, and keeps it private from the servers. Apply this random permutation to reorder the message vectors and define $\tilde{\W}[t]\defeq \W[\pi(t)]$ for $t\in\{1,\dots,L\}$. In addition, choose $N-1$ distinct $\alpha_1, \alpha_2, \ldots \alpha_{N-1} \in GF(q)$ and consider a $(N-1)\times(N-1)$ Vandermonde matrix as

\[
V=\begin{bmatrix}
    1 & \alpha_{1} & \dots  & \alpha_{1}^{N-2} \\
    1 & \alpha_2  & \dots  & \alpha_{2}^{N-2} \\
    \vdots & \vdots & \ddots & \vdots \\
    1 & \alpha_{N-1}  & \dots  & \alpha_{N-1}^{N-2}
\end{bmatrix}
\]

Also, consider $\pi'$ as a random permutation of the set $\{1,2,\dots,N-1\}$ and apply this random permutation to the columns of $V$ to get $\tilde{V}=[\tilde{\alpha}_{i,j}]_{(N-1)\times(N-1)}$.

\textbf{Step 2:} For each $(\vv_m(i_1),\dots,\vv_m(i_{N-1}))\in\mathcal{V}_N\setminus\mathcal{V}_N^{(\theta)}$, $m=0,\dots,(q^K-1)^{N-1}-(q-1)^{N-1}-1$, repeat the following:

\begin{enumerate}[(i)]
  \item User asks server~1 to send back
  \begin{equation}\label{eq:step3-1}
    \sum\limits_{j=1}^{N-1}\vv_m^T(i_j)\tilde{\W}[j+m(N-1)].
  \end{equation}

  \item User asks server~$n$, $n=2,\dots,N$ to send back
  \begin{equation}\label{eq:step3-2}
    \sum\limits_{j=1}^{N-1}(\vv_m^T(i_j)+\tilde{\alpha}_{j,n-1}\vv^T(\theta))\tilde{\W}[j+m(N-1)].
  \end{equation}where $\tilde{\alpha}_{j,n-1}$ is the element of permuted Vandermonde matrix $\tilde{V}$ in the $j$-th row and the $(n-1)$-th column.
\end{enumerate}

We show in the proof of correctness that in each round of this step (i.e., for each $m$), the desired combination $\vv^T(\theta)\tilde{\W}[t]$ is retrieved over $N-1$ message layers.

\textbf{Step 3:} For each $(\vv_m(i_1),\dots,\vv_m(i_{N-1}))\in\mathcal{V}_N^{(\theta)}$, $m=0,\dots,(q-1)^{N-1}-1$ repeat the following:
\begin{enumerate}[(i)]
  \item User asks server~1 to send back
  \begin{equation}\label{eq:step4-1}
    \sum\limits_{j=1}^{N-1}\vv_m^T(i_j)\tilde{\W}[j+(q^K-1)^{N-1}(N-1)+mN].
  \end{equation}

  \item User asks server~$n$, $n=2,\dots,N-1$ to send back
  \begin{equation}\label{eq:step4-2}
    \sum\limits_{j=1}^{N-1}\tilde{\alpha}_{j,n}\vv_m^T(i_j)\tilde{\W}[j+(q^K-1)^{N-1}(N-1)+mN].
  \end{equation}where $\tilde{\alpha}_{j,n}$ is the element of permuted Vandermonde matrix $\tilde{V}$ in the $j$-th row and the $n$-th column.

  \item User asks server~$N$ to send back
  \begin{equation}\label{eq:step4-3}
    \vv_m^T(i_{N-1})\tilde{\W}[N+(q^K-1)^{N-1}(N-1)+mN]+\sum\limits_{j=1}^{N-2}\vv_m^T(i_j)\tilde{\W}[j+(q^K-1)^{N-1}(N-1)+mN].
  \end{equation}

\end{enumerate}

Note that the set of requests to each server are sent in a random order.

We show in the proof of correctness that in each round of this step (i.e., for each $m$), the desired combination $\vv^T(\theta)\tilde{\W}[t]$ is retrieved over $N$ message layers.

\subsection{Proof of correctness}\label{subsec:correct}
To prove the correctness, we show that the user can recover $\{\vv^T(\theta)\W[t]\}_{t=1}^L$ from the combinations \eqref{eq:step3-1}-\eqref{eq:step4-3}, received from $N$ servers, while the rate of the scheme is equal to \eqref{eq:ach_gen}.

In Step~2, we have $|\mathcal{V}_N\setminus\mathcal{V}_N^{(\theta)}|=(q^K-1)^{N-1}-(q-1)^{N-1}$ rounds. In each round $m$, $m\in \{0,\dots,(q^K-1)^{N-1}-(q-1)^{N-1}-1$\}, $N-1$ layers of the desired combination are recovered. The reason follows.
Subtracting \eqref{eq:step3-1} from \eqref{eq:step3-2}, the user has access to
\begin{equation}\label{eq:correct}
    \sum\limits_{j=1}^{N-1}\tilde{\alpha}_{j,n-1}\vv^T(\theta)\tilde{\W}[j+m(N-1)],
\end{equation}
for $n=2,\dots,N$. Since Vandemonde matrix is full rank, \eqref{eq:correct} provides $N-1$ independent linear combinations of $\vv^T(\theta)\tilde{\W}[1+m(N-1)],\dots,\vv^T(\theta)\tilde{\W}[N-1+m(N-1)]$ (that is $N-1$ layers of the desired combination). Thus, we obtain $\vv^T(\theta)\tilde{\W}[1+m(N-1)],\dots,\vv^T(\theta)\tilde{\W}[N-1+m(N-1)]$ from \eqref{eq:correct}. In total,
\begin{equation}\label{eq:L_step3}
    (N-1)((q^K-1)^{N-1}-(q-1)^{N-1})
\end{equation}
layers of the desired combination are recovered in Step~2. The user downloads one equation from each server in each round. Thus, the total number of the downloaded equations in this step is
\begin{equation}\label{eq:Q_step3}
    N((q^K-1)^{N-1}-(q-1)^{N-1}).
\end{equation}

In Step~3, we have $|\mathcal{V}_N^{(\theta)}|=(q-1)^{N-1}$ rounds. In each round $m$,  $m \in \{0,\dots,(q-1)^{N-1}-1\}$,  $N$ layers of the desired combination are recovered. The reason follows.
Since Vandemonde matrix is full rank, the user from \eqref{eq:step4-1} and \eqref{eq:step4-2} has access to the $N-1$ independent linear combinations of
$$\vv_m^T(i_1)\tilde{\W}[1+(q^K-1)^{N-1}(N-1)+mN],\dots,\vv_m^T(i_{N-1})\tilde{\W}[N-1+(q^K-1)^{N-1}(N-1)+mN].$$
These are $N-1$ layers of the desired combination, as in this step, the coefficient vectors satisfy $(\vv_m(i_1),\dots,\vv_m(i_{N-1}))\in\mathcal{V}_N^{(\theta)}$, and thus  all are parallel to $\vv(\theta)$.
Eliminating the $N-2$ layers from \eqref{eq:step4-3}, the user recovers the $N$-th layer in this step which is $\vv_m^T(i_{N-1})\tilde{\W}[N+(q^K-1)^{N-1}(N-1)+mN]$. Therefore, $N$ layers of the desired combination are recovered for each $m$.

In total,
\begin{equation}\label{eq:L_step4}
    N(q-1)^{N-1}
\end{equation}
layers of the desired combination are recovered in Step~2. The user downloads one equation from each server in each round. Thus, the total number of the downloaded equations in this step is
\begin{equation}\label{eq:Q_step4}
    N(q-1)^{N-1}.
\end{equation}

From \eqref{eq:Q_step3} and \eqref{eq:Q_step4}, the total number of downloads is
\begin{equation*}
    Q=N(q^K-1)^{N-1}
\end{equation*}
and totally
\begin{equation}\label{eq:L}
    L=(N-1)(q^K-1)^{N-1}+(q-1)^{N-1}
\end{equation}
layers of the desired combination are recovered and so the rate of the code is as \eqref{eq:ach_gen}.
\subsection{Proof of privacy}\label{subsec:privacy}
The privacy proof is based on the fact that we preserve the equal number of requests for any possible combination vector in addition to using a random permutation over the message layers. In addition, the requests to each servers are sent in random order.

First, consider server~1 with its requests \eqref{eq:step3-1} and \eqref{eq:step4-1}. As seen, server~1 only observes that the user requests a linear combination of $N-1$ message layers with all possible coefficient vectors. The indices of the message layers do not leak any information about the requested combinations vector $\vv(\theta)$, thanks to the random permutation of the message layers. In addition, due to the random order of the requests, asking for all possible coefficient vectors makes $Q_{\vv(i)}^{(1)},A_{\vv(i)}^{(1)}$ equiprobable for $i=1,\dots,q^K-1$.

Now, consider server~$n$, $n=2,\dots,N$, with its requests \eqref{eq:step3-2}, \eqref{eq:step4-2} and \eqref{eq:step4-3}. Again, server~$n$ only observes that the user requests a linear combination of $N-1$ message layers with all possible coefficient vectors (with a random order). Because:
\begin{itemize}
  \item The set of $\mathcal{\bar{V}}_m=\{(\vv_m(i_1),\dots,\vv_m(i_{N-1}))\}$ that is used in the scheme covers $\mathcal{V}_N$ (i.e., all possible $(N-1)$-tuples of vectors $(\vv(i_1),\dots,\vv(i_{N-1}))$ with each element from $\mathcal{V}$).
  \item The set $\{(\vv_m(i_1)+\tilde{\alpha}_{1,n-1}\vv(\theta),\dots,\vv_m(i_{N-1})+\tilde{\alpha}_{N-1,n-1}\vv(\theta))\}$ is equal to $\mathcal{\bar{V}}_m$.
  \item When $(\vv_m(i_1),\dots,\vv_m(i_{N-1}))\in\mathcal{V}_N^{(\theta)}$ (in Step~3), two sets $\{(\vv_m(i_1)+\tilde{\alpha}_{1,n-1}\vv(\theta),\dots,\vv_m(i_{N-1})+\tilde{\alpha}_{N-1,n-1}\vv(\theta))\}$ and $\{(\tilde{\alpha}_{1,n}\vv_m(i_1),\dots,\tilde{\alpha}_{N-1,n}\vv_m(i_{N-1}))\}$ are equal.
\end{itemize}

Thus, it can be shown that for any combination vector $\vv(j),j=2,\dots,\tilde{K}$, there is a random permutation $\pi$ of the set $\{1,2,\dots,L\}$ that maps the request of $\vv(\theta)$ to the request of $\vv(j)$.


\vfill

\end{document}